\documentclass[conference]{IEEEtran}
\usepackage{balance}
\usepackage{algorithm}
\usepackage{algorithmicx}
\usepackage{algpseudocode}
\usepackage{multirow}
\usepackage{multicol}
\usepackage{color}
\usepackage{epsfig}
\usepackage{url}
\usepackage{subfigure}
\usepackage{graphics}
\usepackage{amsmath}
\usepackage{amssymb}
\usepackage{comment}
\usepackage{makeidx}  
\usepackage{comment}
\usepackage{xr}
\usepackage{epstopdf}
\usepackage{float}
\usepackage{dblfloatfix}

\usepackage[normalem]{ulem}

\algnewcommand{\LineComment}[1]{\State \(\triangleright\) #1}

\def\BibTeX{{\rm B\kern-.05em{\sc i\kern-.025em b}\kern-.08em
    T\kern-.1667em\lower.7ex\hbox{E}\kern-.125emX}}
\begin{document}

\title{Clustering based opcode graph generation for malware variant detection} 

\author{\IEEEauthorblockN{1\textsuperscript{st} Fok Kar Wai}
\IEEEauthorblockA{\textit{Cybersecurity Strategic Technology Centre} \\
\textit{ST Engineering}\\
Singapore, Singapore \\
fok.karwai@stengg.com}
\and
\IEEEauthorblockN{2\textsuperscript{nd} Vrizlynn L. L. Thing}
\IEEEauthorblockA{\textit{Cybersecurity Strategic Technology Centre} \\
\textit{ST Engineering}\\
Singapore, Singapore \\
vriz@ieee.org}
}
\maketitle

\begin{abstract} 

Malwares are the key means leveraged by threat actors in the cyber space for their attacks.
    There is a large array of commercial solutions in the market and significant scientific research
    to tackle the challenge of the detection and defense against malwares. At the same time,
    attackers also advance their capabilities in creating polymorphic and metamorphic malwares
    to make it increasingly challenging for existing solutions. To tackle this issue,
    we propose a methodology to perform malware detection and family attribution. The proposed
    methodology first performs the extraction of opcodes from malwares in each family 
    and constructs their respective opcode graphs. We explore the use of clustering algorithms
    on the opcode graphs to detect clusters of malwares within the same malware family.
    Such clusters can be seen as belonging to different sub-family groups.
    Opcode graph signatures are built from each detected cluster. Hence, for each malware family,
    a group of signatures is generated to represent the family.
    These signatures are used to classify an unknown sample as benign or belonging 
    to one the malware families. We evaluate our methodology by performing experiments
    on a dataset consisting of both benign files and malware samples belonging to
    a number of different malware families and comparing the results to existing approach.

\end{abstract}



\section{Introduction}

Malware is often used as a means to gain an initial compromise of victim hosts
commonly through activities such as social engineering and phishing schemes. Once targets are
infected, a variety of malicious activities are furthered performed by the malwares. These activities include
lateral movement to infect more targets in the same network, exfiltration or destruction of 
valuable data and the disruption of critical systems. 
The detection of malwares is therefore a must for protecting critical systems and hence is a widely researched
topic. There is a large variety of research works and methodologies on the detection of malwares 
which have been introduced by various surveys \cite{survey-2007, survey-heuristic-2013,
survey-malware-2014, survey-data-mining-2017}. However, it has become increasingly challenging
for these methods to have ideal detection performances due to the rapidly evolving malwares and growing
pool of different malware families \cite{fireeye-report-2020}.
In the SonicWall's 2021 Cyber Threat Report
\cite{sonicwall-sec-report-2021},
their solutions found a total of 589,313 new malware variants in 2020. Overall, 74\% more
never-before-seen malware variants were identified in 2020 than in 2019.

Malware authors apply various types of techniques on the creation of malware variants
in order to evade detection from the variety of defense mechanisms. 
Such obfuscation techniques include encryption, polymorphism and metamorphism which have been 
introduced in research works such as in \cite{obfuscation-survey-2010}. 
In particular, metamorphism provides the capability to alter the internal structure of each malware, whilst being
able to maintain the same functional outputs \cite{meta-hunting-2011, meta-morphing-2013, 
meta-entropy-2013}. 
Thus it is possible to produce an infinite pool of malwares within the same family. These malwares achieve the same
purpose but are not identical. Common signature-based
solutions therefore become ineffective as each of these malwares have different signatures. 
This also makes it challenging for other types of methods such as dynamic and behaviour analysis
to effectively detect these continuously morphing malwares that have evasive mechanisms during runtime.


Static analysis, which involves analyzing the file properties and structure, may also be 
relied upon to detect obfuscated malware. In this paper, our proposed methodology focuses on the
analysis of one such static property, which is opcodes. Opcodes are machine instruction codes that define 
the operations that needs to be performed when the file is executed. Hence, it is a 
representation of a file's structure and operations and is useful for malware detection. 


In this work, we extract opcodes from malware samples, build opcode graphs using bigrams and 
perform graph similarity comparison for malware classification and family attribution.
However differing from the above previous works and graph-based methods in particular, 
we integrate the use of clustering algorithms on the graphs belonging to 
samples within the same families. The purpose of this idea is to first detect potential clusters of
sub-family samples as a preliminary step, as malwares belonging
to the same family may be generated from different versions of the family codebase. Hence 
there may be some considerable differences in the structure of the malwares generated by different
code versions within the same family. It is then meaningful to detect these different groups to be 
able to boost detection accuracy. Opcode bigrams belonging to clustered samples are used to build
a single opcode graph which represents its signature. Each malware family is thus represented by a 
set of opcode graph signatures, with each signature belonging to a different sub-family or version.
For an incoming unknown sample, its opgraph is extracted and compared against each signature to
find the closest matching family for classification.

In summary, these are the contributions of our paper:
\begin{enumerate}
    \item An opcode-based approach for malware detection and attribution is proposed.
        The method builds opcode graph signatures for each malware family, which can be used for
        similarity comparison against samples and classify them to one of the malware families.
    \item Clustering algorithms are explored to perform the detection of clusters on samples
        belonging to the same malware family. The existence of different clusters of samples
        that have considerable difference in characteristics and therefore potentially
        belong to separate sub-families is demonstrated as a result.
    \item The proposed methodology is evaluated on a dataset comprising of 2300 malware samples
        belonging to six different malware families and 580 benign file samples. Both binary
        classification experiments for general malware detection and multi-class classification
        experiments for family attribution are performed.
\end{enumerate}

In the following section, we discuss the related works. In Section~\ref{sec:method}, we present
our methodology in detail and describe its various steps. Experiments and performance evaluation
of the proposed method are presented in section~\ref{sec:eval}. Finally, future improvements and
conclusions are discussed in section~\ref{sec:conclusion}


\section{Related works}
\label{sec:related}

Vast research has been carried out in the areas of malware analysis due to the rampant growth
of the issue and the critical need for defense mechanisms against malwares. 
Many research surveys have been published to provide the community 
with a comprehensive breakdown of features and techniques explored. \cite{survey-ml-2017, 
survey-data-mining-2017} are some recent surveys. Malware analysis can be broadly categorised into two
categories, static analysis and dynamic analysis. In static analysis, extraction of features based
on the content of the files without executing them is performed. Such features include file strings, byte sequences, 
opcodes, static API calls and control flow graphs. 
There are also features unique to different types of applications such as android. In \cite{android-attribution},
n-gram analysis techniques were applied on XML and DEX files found within the application's 
Android Package Kit (APK).
On the other hand, dynamic analysis involve
observing the behavior of malwares during execution in a simulated environment. This approach allows more
information related to the malwares' interactions, such as its dynamic API calls,
network activity, access and modifications to the file systems and registries to be collected. 
In this paper we focus on the static analysis feature of opcodes for malware family attribution and 
detection of metamorphic malware.
Santos et al. \cite{opcode-freq-fam-2010} represented each malware
sample as a vector containing the frequencies of fixed length opcode sequences. Similarity 
comparison is performed against variants belonging to a set of different malware families and
attribute the sample to the closest family. 
Rad and Masrom \cite{opcode-histo-2011} extracted opcode frequencies as histograms and applies
histogram dissimilarity metrics to detect similarities to
metamorphic families and hence classify them.
Toderici and Stamp \cite{chi-square-2012} proposed to combine chi-squared statistical test with
HMM method.
Each program is represented by a spectrum of opcode frequencies and chi-squared test is performed
to determine whether it matches the expected spectrum of frequencies from a malware family.
This method is effective at metamorphic virus detection but performs poorly when 
techniques such as copying benign codes into malware are used. 
Shangmugam et al. \cite{cipher-2013} applied the use of a simple substitution cipher technique as
a measure of similarity between the opcode distribution matrix of each metamorphic family to that
of an unknown sample.
Fazlali et al. \cite{opcode-freq-meta-2016} extracted 
an opcode feature set such as normalized frequency count of opcodes and appearance of high 
frequency opcodes. The feature set is used to train models of six different decision tree based
machine learning algorithms such as Random forest to classify samples into metamorphic families or
as benign.
Sahay and Sharma \cite{opcode-freq-kmeans-2016} investigated and found that malwares generated from the
same metamorphic virus kit are similar in size. Thus they applied K-Means clustering on their size
values to detect clusters of metamorphic malwares from a dataset. Opcode frequency based features
are then extracted from each cluster and used to train common classifiers for detection. 
Okane et al. \cite{reduced-opcode-2016} explored the use of runtime opcodes during the execution
of the program for obfuscated malware such as encrypted malware. 
Support vector machine (SVM) was applied on density histogram of runtime opcodes
as features. 

Recently, works have applied deep learning on opcode-based features. Darem et al.~\cite{dl-visual}
uses the popular method of converting features into images.
Opcode n-gram features were converted into malware binary code and finally into gray-scale images
before feeding into a Convolutional Neural Network (CNN) model. Their final model is an ensemble
of CNN and XGBoost, combining both deep learning and traditional machine learning.~\cite{dl-byte}
is another image based deep learning work that combines opcode features from both ASM and Bytes
file of the malwares.

Graph-based techniques have also been applied on opcodes for malware analysis. Many of such works extract
weighted directed graph of opcode digrams for each malware sample.
One aspect of such works involve the use of sub-graph matching techniques. Khalilian et al. \cite{g3md-2018}
constructs the opcode graph of samples and applies graph mining techniques to detect frequently occurring
subgraphs across samples of the same malware family. These subgraphs represent micro-signatures
of each family and are used as features to train classifiers for detection. \cite{cfg-subgraph-2014}
is a closely similar work but uses control flow graphs (CFG) instead of opcodes for frequent
sub-graph mining.
Gulmez et al.~\cite{subgraph-histo} performs further engineering of the sub-graphs of opcodes into histograms
containing the degree of each graph node to produce final features as input to the machine learning models.
Alam et al. \cite{cfg-opcode-2015} proposes two different techniques, with one
using CFG and the other using opcodes. One key novelty of their work involves representing both CFG
and opcodes as higher level patterns. This CFG is termed 
as Annotated CFG (ACFG) and they apply sub-graph isomorphism technique for malware pattern matching.
Their second technique performs analysis of pattern distributions by applying sliding windows
on the higher level opcode representations.

The other aspect of graph-based works consist of graph similarity analysis approaches which are 
closely related to our proposed methodology. Runwal et al. \cite{opcode-graph-2012} first extracts the
opcode graph of each malware sample. In order to determine the similarity between two graphs 
extracted from different samples, they proposed a formula to compute a score value from the graph
matrices. They determined a threshold for the score value by performing a comparison of all samples
in each metamorphic family and benign files. To classify new samples, comparison is performed on the
opcode graph of the sample and any sample belonging to each metamorphic family or benign files and the sample
is assigned to the class in which the score falls below the threshold. Kakisim et al. \cite{co-opcode-graphs-2020}
proposes two graph similarity based techniques. Their first technique, coined as Co-opcode Graph
Similarity based Metamorphic Malware Identification method (CGS-MMI), involves
building a single opcode graph to represent each metamorphic family. A similarity comparison of 
opcode graphs of an unknown sample and that of each metamorphic family is done to determine the family
with highest similarity for classification. Their similarity score is computed using a 2-D 
correlation coefficient formula which uses the mean of the weights of edges in the graphs. 
Their second proposed technique, Higher-Level Engine Signatures based Metamorphic Malware Identification
method (HLES-MMI), aims to extract a higher level signature from the 
representative opcode graph of each metamorphic family. The idea is to only retain nodes and edges
with the largest weights in each graph which indicates a specific pattern signature for each family.
A metric is then used to determine the similarity among signatures for classification.

As our work also involves the use of clustering algorithms, following is a discussion on some
clustering works. Zhang et al. \cite{cluster-ensemble-2017} uses both static and dynamic features
and applies clustering algorithms such as K-means and Hierarchical clustering. An ensemble 
method to combine both algorithms and improve robustness of results is proposed. Pitolli et al.
\cite{malfamaware-2020} also use hybrid features but apply BIRCH clustering. Clustering is performed
on the input dataset to group samples for the purpose of identifying the set of families that exist
and attribute the samples to those families.

Hu et al. \cite{mutantx-2013} propose a clustering-based work using opcode n-grams. Feature vectors
containing occurrences of each opcode n-gram are used as input to calculate the Euclidean distance
and hence determine program similarities. The authors argue that classic algorithms such as K-means
and hierarchical clustering do not scale and propose the use of a prototype-based clustering
method to cluster samples in their dataset of 20 malware families.
Wang et al. \cite{opcode-clustering-2016} extracts opcode bigrams and uses information
entropy on the bigram probabilities to select a smaller set of important bigrams as features.
They propose a Fast Density-based Clustering which is an improvement over the standard density-based
algorithm.

Differing from the above clustering works, our proposed methodology does not apply clustering
on a dataset as a whole to identify the malware families or use clustering as the main means to 
classify new samples to their families. 
Instead we apply clustering separately on samples belonging to each malware family using a 
labelled dataset of samples with family categorisation. The purpose of this to attain a more fine-grained
attribution capability by detecting potential sub-family groups that may have considerable
differences even if they belong to the same family. In \cite{mutantx-2013}, it was reported that
previous clustering approaches resulted in families being separated into several sub-family clusters, more 
specifically the detection of 50 clusters from a dataset of 20 families, due to highly diverse
samples. This is an indication that our approach would be a meaningful research direction.

Our clustering approach is integrated to the graph similarity analysis techniques in \cite{opcode-graph-2012,
co-opcode-graphs-2020}. We follow these methods of similarity comparison 
as they have been shown to be effective in the detection of malware samples belonging to different
metamorphic families. 
However, instead of applying the similarity comparison directly on samples, 
our approach applies such techniques as a means for similarity and distance calculations for input
to the clustering algorithm to detect sub-family clusters within the same family. 

\section{Methodology}
\label{sec:method}

\begin{figure*}[t]
       \centering{
           \includegraphics[width=\textwidth]{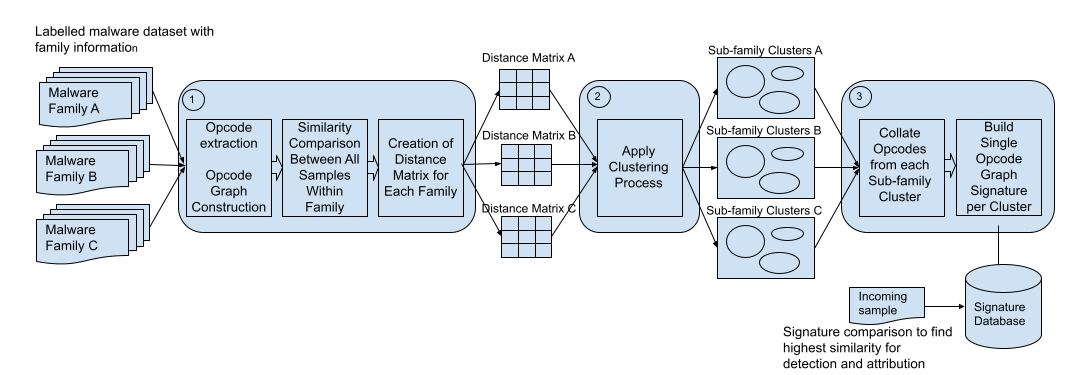}
            \caption{{Three-stage process of methodology}}
                \label{fig:process}
                   }
          \vspace{-0.2cm}
\end{figure*}

Our proposed methodology can be organised as a three-stage process consisting of the following:
\begin{enumerate}
    \item Creation of family distance matrix via similarity comparison of opcode graphs
    \item Detection of sub-family groups via application of clustering algorithm
    \item Building of opcode graph signatures from clusters for detection and family attribution
\end{enumerate}

An illustration of the above process is shown in Fig.~\ref{fig:process}.

\subsection{Stage 1: Creation of family distance matrix}
In order to perform clustering on samples belonging to each malware family and detect sub-family
clusters, the first step is to create the feature inputs. Our inputs to the clustering phase are 
a set of $N \times N$ matrices, where $N$ represents the total number of samples belonging to each 
malware family in the dataset. Each matrix contains the distance values of
each sample in the family to all the other samples within the same family.

\subsubsection{Opcode graph construction}
\label{subsec:graph-construction}
In order to perform the similarity comparison to create the distance matrices, we leverage on 
opcode graph similarity comparison techniques. Specifically for opcode graph construction, we 
choose to use the weighted directed graphs in \cite{opcode-graph-2012, g3md-2018}.

All file samples in our dataset are disassembled to obtain their opcode sequences.
From the set of opcode sequences, two key information are extracted. First is the list of all unique opcodes
found. The other is the set of opcode bigrams which are pairs of opcodes in which one in the pair 
follows the other subsequently in the sequence. Using the above information, a weighted directed 
graph of opcodes is constructed for each sample in the dataset. Each node in the graph represents
a unique opcode in the dataset. The edge from one node to another represents an opcode bigram (\textit{i, j}),
where \textit{j} occurs subsequently after \textit{i} in the opcode sequence of the sample. To
compute an edge weight for (\textit{i, j}), we first total the number of occurrences for (\textit{i, j})
in the sample. This value is then divided by the total number of opcode bigrams where \textit{i}
belongs to the first member in the pair. The result is an edge weight representing the 
probability that \textit{j} occurs after \textit{i}, whenever \textit{i} appears in the 
sample's opcode sequence.

\vspace{1cm}
\begin{figure}[!htbp]
\centering
  \subfigure[Graph for Sequence 1]{
  \centering{
  \includegraphics[scale=0.33]{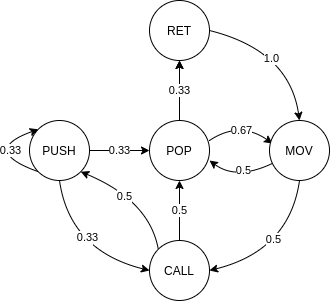}
  \label{subfig:graph1}}}
  \subfigure[Graph for Sequence 2]{
  \centering{
  \includegraphics[scale=0.33]{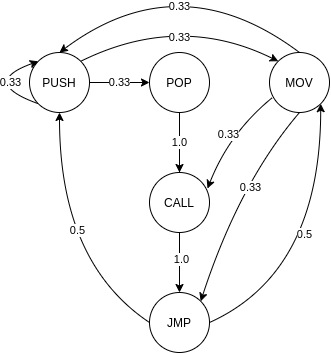}
  \label{subfig:graph2}}}
  {\caption{Constructed opgraph graphs for sequences in Table~\ref{tab:seq}}\label{fig:opcode-graphs}} 
  \vspace{-0.3cm}
\end{figure}

\begin{table}[]
    \centering
\begin{tabular}{cc}
Sequence 1 & Sequence 2 \\ \hline
PUSH       & MOV        \\
POP        & CALL       \\
MOV        & JMP        \\
POP        & MOV        \\
RET        & PUSH       \\
MOV        & PUSH       \\
CALL       & POP        \\
PUSH       & CALL       \\
PUSH       & JMP        \\
CALL       & PUSH       \\
POP        & MOV        \\
MOV        & JMP        \\ \hline
\end{tabular}
    \vspace{0.1cm}
\caption{\label{tab:seq} Example illustration of two opcode sequences}
\end{table}

Table~\ref{tab:seq} and Fig. \ref{fig:opcode-graphs} provide an example illustration of the 
opcode graph construction from opcode sequences. We consider a simple example scenario where the input 
contains only two file samples. Table~\ref{tab:seq} shows the opcode sequences of each sample. 
We may then refer to Fig. \ref{fig:opcode-graphs} to observe the opcode 
graphs that are constructed from each of the sequences, noting that the edges in the graphs
contain the probability value of one specific opcode appearing after another specific opcode
based on the direction of the edge.
For example, in Fig. \ref{subfig:graph1}, we can observe that after a PUSH instruction, either a
POP, CALL or another PUSH appears subsequently each with an equal probability.

\subsubsection{Opcode filtering}
During the opcode sequence and opcode bigram extraction process, we had also performed filtering
of opcodes to remove some from consideration in the opcode graph construction. As previously defined,
the number of nodes in the opcode graph is the number of unique opcodes found in the samples of the
entire dataset. Thus depending on how many unique opcodes there are, the size of the opcode graph 
may become very large. We filter and remove rarely occurring opcodes which lead to the graphs 
becoming unnecessarily large while not providing significant information or negatively impacting
the similarity comparisons. 
More specifically, the process involves sorting the opcode bigrams
according to their number of occurrences and we calculate the total number of bigrams.
A simple threshold percentage is then set and applied on this total value. If the threshold value is
90\%, the top occurring opcode bigrams that contribute to 90\% of the total number of bigrams is
retained while the remaining ones are filtered off. From the pool of filtered bigrams, unique opcodes
that no longer exist in any of the retained bigrams are removed from consideration thereby
reducing the size of the opcode graphs.

\subsubsection{Similarity comparison of opcode graphs}
\label{subsec:distance}
After the opcode graph construction for the samples, we proceed to perform similarity comparison of all
samples within the same malware family in the dataset. The measure of similarity between samples, 
i.e. distance matrix, is a possible feature input to clustering algorithms for them to able to group our samples within
the same family into different sub-family clusters.

To compute a similarity score between any two samples using their opcode graphs, we refer to
the computation formula proposed in \cite{opcode-graph-2012}. The scoring function is simple and 
efficient and is able to sufficiently reflect the similarity in the structure of the opcode graphs.
Recall that each opcode graph is an $N \times N$ weighted directed graph.
Hence the more similar one sample is to another sample, their edge weights
should be closer in value in general.
The scoring approach takes the above into consideration. 
It computes the difference in values of the corresponding edge weights in the two samples. 
Then the final score is computed using a formula which includes the 
summation of this difference in values between all the edge weights.
Hence, the lower the score value, the more similar any two samples are to one another. 
A minimum score value of 0 then indicates that the two graphs are identical.
In strict terms, this is in fact a measure of distance as opposed to similarity.

For each malware family, we proceed to perform the above computation to calculate the score between
each sample and all the other samples in the family. Thus, the final output is an $N \times N$
distance matrix for each malware family. Each matrix will be separately used as a feature input for clustering
in the next stage.

\subsection{Stage 2: Detection of sub-family groups via application of clustering algorithms}
\label{subsec:clustering}
To perform the clustering in this stage, we need to first select a clustering algorithm capable of
meeting the requirements for our problem. An important criteria for the algorithm is to be able to
make a decision by itself on the number of clusters there are. Our input dataset is labelled
with specific number of families. However our aim is to perform clustering within each family to
detect the potential sub-family groups which is unknown and meant to be discovered through this
process. Hence clustering algorithms such as the K-means are not suitable as they require the
number of clusters to be defined. A likely candidate for this problem is the Density-Based Spatial 
Clustering of Applications with Noise (DBSCAN) algorithm \cite{dbscan-1996} which is able to
automatically discover the number of clusters. Density-based clustering
methods like DBSCAN do not group every point in the input into a cluster. Instead, only points that
are very tightly packed are clustered and any points that are far from clusters and do not have their
own close neighbours are considered as noise. This is applicable for the input scenario, where 
not all samples may belong to a group of samples generated by the same sub-family version. There may
be lone samples generated by new versions which are not yet commonly found, and such samples
could be detected as noise in the DBSCAN clustering process.

\subsubsection{Investigation of DBSCAN parameter}
The DBSCAN algorithm automatically discovers the number of clusters based on a very important input
parameter, \textit{eps}. This parameter indicates the maximum distance between two samples for them
to be considered in the same neighborhood for clustering. 

\begin{figure}[!htbp]
\centering
  \subfigure[Clustering results using \textit{eps}=0.01]{
  \centering{
  \includegraphics[scale=0.52]{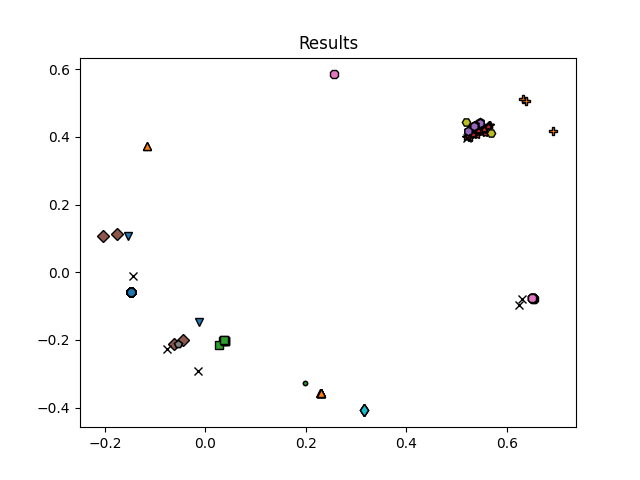}
  \label{subfig:q-graph1}}}
  \subfigure[Subplot of the graph in \ref{subfig:q-graph1}]{
  \centering{
  \includegraphics[scale=0.72]{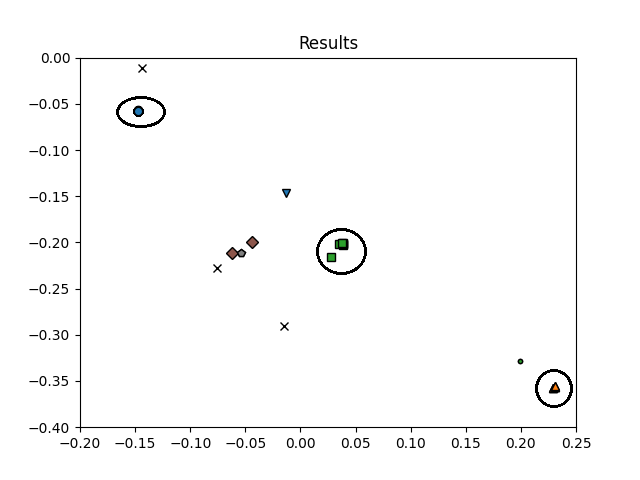}
  \label{subfig:q-graph2}}}
  {\caption{DBSCAN clustering results on Qakbot malware samples}\label{fig:quakbot-graphs}} 
  \vspace{-0.3cm}
\end{figure}

As the selection of this parameter is 
critical, we performed an investigation on how it may affect the outcome of clustering for the 
different malware families in our dataset with relation to the distance function used and distance 
matrices produced in Section~\ref{subsec:distance}
. Fig~\ref{fig:quakbot-graphs} shows clustering results
for Qakbot malware samples. In particular Fig~\ref{subfig:q-graph1} shows the overall clustering
when we used an \textit{eps} value of 0.01. 
Clusters are indicated by the different shape patterns in the graph. Points that have the ``x''
pattern are noise points that do not belong to any clusters.
Based on the graphical view, our observation
is that the clustering results were promising. To better demonstrate, Fig~\ref{subfig:q-graph2} 
shows a subplot of ~\ref{subfig:q-graph1} and the clusters being marked out in the graph contain a considerable
number of samples with very high similarity.

\begin{figure}[!htbp]
\centering
  \subfigure[Clustering results using \textit{eps}=0.01]{
  \centering{
  \includegraphics[scale=0.52]{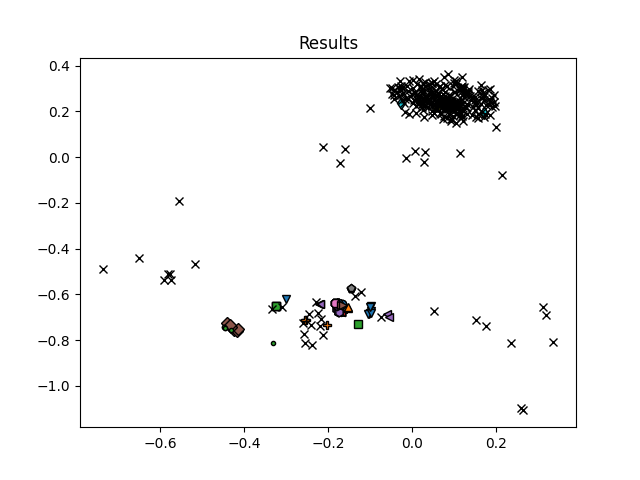}
  \label{subfig:f-graph1}}}
  \subfigure[Clustering results using \textit{eps}=0.1]{
  \centering{
  \includegraphics[scale=0.52]{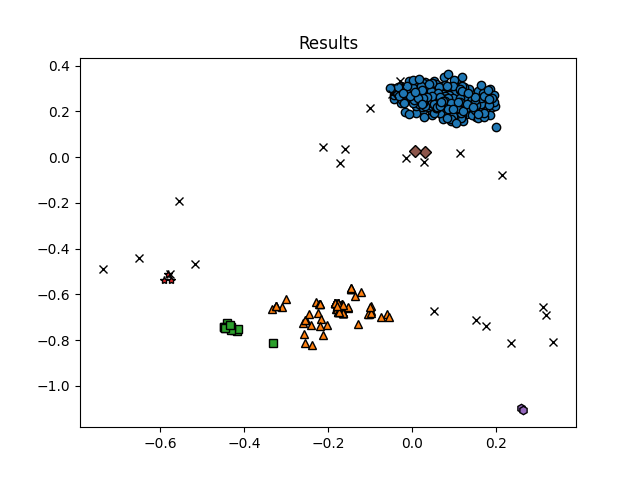}
  \label{subfig:f-graph2}}}
  {\caption{DBSCAN clustering results on Formbook malware samples}\label{fig:formbook-graphs}} 
  \vspace{-0.3cm}
\end{figure}

Fig~\ref{fig:formbook-graphs} shows the clustering results on the Formbook malware samples using
two different \textit{eps} values. We can observe in Fig~\ref{subfig:f-graph1} when the \textit{eps}
value is 0.01, the same as that in the Qakbot investigation, DBSCAN fails to cluster most of the 
samples which are all detected as noise. When a larger \textit{eps} value of 0.1 is used as shown
in Fig~\ref{subfig:f-graph2}, we can observe that most samples are clustered. This provides a 
conclusion that as each malware family has its own unique properties, clustering cannot be applied
exactly the same way for all of them. There may be a need to apply different \textit{eps} values for
different circumstances.

\subsubsection{Clustering process}
Based on the above conclusion, in order to obtain optimal clustering results for different malware family,
we came up with a clustering process that involves multiple rounds of DBSCAN clustering on a
set of \textit{eps} values. We first manually define the \textit{eps} values and sort them
according to increasing order. DBSCAN clustering is then applied on the input samples using the smallest
\textit{eps} value. All points detected as noise are then retained and sent for another round of 
clustering using the following \textit{eps} value.

    \begin{algorithm*}
    \caption{Malware sub-family clustering process}
    \begin{algorithmic}[1]

        \Procedure{PerformClustering}{}
        \State $M \gets \{M^1, ..., M^N\}$ \Comment{Set of distance matrices for N families}
        \State $S \gets \{S1, ..., S^N\}$ \Comment{Corresponding samples for each distance matrix}
        \State $C \gets \{\}$ \Comment{Variable to store final result of detected clusters}
        \State $eps \gets \{eps^1,...,eps^D\}$ \Comment{Define set of \textit{eps} values}
        \For {$n\gets 1$ to $N$}
            \State $m \gets M^n$
            \State $s \gets S^n$
            \For {$d\gets 1$ to $D$}
                \State $labels \gets \Call{DbscanClustering}{eps^d, m}$
                    \Comment{Perform DBSCAN clustering with selected \textit{eps}}
                \State $clusters, noise \gets \Call{ProcessLabels}{s, labels}$
                    \Comment{Extract clustered/noise samples from DBSCAN labels}
                \State $m \gets \Call{RecomputeDistanceMatrix}{m, noise}$
                    \Comment{Compute new matrix using noise samples for next round}
                \State $C \gets C + clusters$ \Comment Add detected clusters to final result
            \EndFor
        \EndFor
        \State \Return C 
            \Comment{Final result containing all detected clusters}
    \EndProcedure

    \end{algorithmic}
    \label{fig:algo}
    \end{algorithm*}

The above clustering process is illustrated in Algorithm~\ref{fig:algo}. The set of \textit{eps} values
with increasing order allows DBSCAN to cater to the distribution of distance values for different
malware families. By starting with a low \textit{eps} value, there is good clustering performance for 
malware families such as Qakbot that contain distinct sub-family groups with high sample similarity
as shown in Fig~\ref{fig:quakbot-graphs}. When the same value is applied for other families such as
Formbook, most samples are detected as noise in Fig~\ref{subfig:f-graph1}. In such cases, when the
algorithm progresses to the next \textit{eps} which has a higher value, we will obtain
better clustering results.

Once the clustering process is complete, we would obtain groups of clustered samples for each malware family.
Each group contains samples that may potentially have been generated from a different version
of the malware code. However at the end of this process, there may still be remaining samples that
have been labelled as noise even after all rounds of clustering. We consider such lone samples
as their own groups containing only a single member.

\bgroup
\def\arraystretch{1.5}
\begin{table}[]
    \begin{tabular}{|c|c|c|c|c|}
        \hline
        \textit{eps Setting}      & Family     & No. Samples & No. Clusters & No. Unclustered \\ \hline
        \multirow{3}{*}{0.01}     & Qakbot     & 320         & 11           & 9               \\ \cline{2-5} 
                                  & AgentTesla & 400         & 13           & 363             \\ \cline{2-5} 
                                                            & Formbook   & 320         & 13           & 252             \\ \hline
                                                            \multirow{3}{*}{0.1}      & Qakbot     & 320         & 7            & 2               \\ \cline{2-5} 
                                                                                      & AgentTesla & 400         & 22           & 98              \\ \cline{2-5} 
                                                                                                                & Formbook   & 320         & 27           & 127             \\ \hline
                                                                                                                \multirow{3}{*}{Proposed} & Qakbot     & 320         & 13           & 2               \\ \cline{2-5} 
                                                                                                                                          & AgentTesla & 400         & 37           & 4               \\ \cline{2-5} 
                                                                                                                                                                    & Formbook   & 320         & 40           & 6               \\ \hline
    \end{tabular}
    \vspace{0.1cm}
    \caption{\label{tab:clustering-evidence} Evaluation of clusters detected over different \textit{eps} settings}
\end{table}
\egroup

We provide an evaluation of our proposed clustering scheme by looking at the concrete number of 
clusters detected and unclustered samples belonging to three malware families in our dataset on different \textit{eps} 
settings.
We can observe in Table~\ref{tab:clustering-evidence} that for an \textit{eps} value setting of 0.01, 
the number of unclustered samples for Qakbot is very low. 
This coincides with our earlier explanation with reference to Fig~\ref{fig:quakbot-graphs},
that such a setting was effective for this particular malware family. On the other hand, the number of
unclustered samples for AgentTesla and Formbook form the majority.
This situation is improved for both families when the setting of 0.1 is used as more samples
are able to be clustered. 
For Qakbot, observe that the number of clusters detected in this setting
is 7 which is lesser than the previous of 11. This indicates that more distinct sub-family clusters
could have been detected which the algorithm was unable to do so in the current setting.
Finally, for our
proposed process of applying increasing \textit{eps} value settings in an iterative fashion,
the clustering results obtained is the best with low number of unclustered samples.
This is an evidence of the benefits of the proposed scheme as compared to directly applying the 
DBSCAN algorithm on a specific \textit{eps} value setting.

\subsection{Stage 3: Opcode Graph Signature Building}
\label{subsec:sig}
After the clustering stage, we proceed to the final stage of building opcode graph signatures which
are used for similarity comparison to detect and classify new samples.

As opposed to building an opgraph graph for each individual malware sample as in Section~\ref{subsec:graph-construction},
a single opcode graph is built for each group of clustered samples formed in the previous stage.
This is done by consolidating the opcode bigrams extracted from every member in the cluster as if
they belonged to a single sample and using them to build a single opcode graph. This graph is
a representation of all members belonging to the cluster as a whole and is thus a signature for the cluster.

Once all signatures have been built, each malware family has a set of opcode graph signatures.
Each signature represents a sub-family group within the malware family. In order to detect a
new sample as malicious and classify it to one of the malware families, the same similarity
comparison approach in Section~\ref{subsec:distance} is used. The incoming sample is disassembled to obtain
its opcode sequence and used to construct its opcode graph. The scoring function is used to
compute the similarity score between the sample's opcode graph and all opcode graph signatures. The
sample is thus classified to the malware family that contains the signature with highest similarity
to the sample.

\section{Performance evaluations}
\label{sec:eval}

\subsection{Dataset}
\label{subsec:dataset}
To evaluate the proposed methodology, we curated a dataset consisting of 2300 malware samples
belonging to six different malware families and 580 benign file samples. Although
the dataset is imbalanced, our methodology does not involve directly applying a 
machine learning technique for learning and prediction. Thus issues
related to imbalanced data in the above approach does not exist in our methodology.

\subsubsection{Malware}
Table~\ref{tab:malware}
shows a breakdown of the malware families and their corresponding sample count in our dataset.
These malwares are collected from the MalwareBazaar project~\cite{bazaar}, a public
malware database with samples contributed by the community. Specifically, our malwares are collected
from a portion of MalwareBazaar's daily lists of malware uploads from the period of November 2020 to
May 2021. In order to use these samples in our experiments, there are two types of verifications that
needed to be performed on the samples. 

Firstly, we need to verify the validity that the collected
samples were indeed malwares. To do so, we leveraged on VirusTotal's~\cite{virustotal} online
file submission API service to perform analysis on the samples. The analysis results contain
verdicts from a list of malware detection engines. To ensure validity, we choose to only use samples
that have been flagged by at least a number of their detection engines (10\% of the engines) in our dataset.

The second action performed was to obtain the malware family information of the collected samples.
As our work's focus is on malware family attribution, the family information needs to be accurate.
While MalwareBazaar provides its own signature information containing malware family labels,
there is a need for further verification. MalwareBazaar results contain information from other
sources such as Hatching Triage malware analysis sandbox~\cite{hatching}. We also leverage
on other sandbox services such as Joe Sandbox~\cite{joe} to obtain analysis data. 
Thus, we only used the malware samples that have the same family information
from multiple of these sources.

\begin{table}[]
\centering
\begin{tabular}{|c|c|}
\hline
Malware Family & Count \\ \hline
AgentTesla     & 500   \\ \hline
Loki           & 411   \\ \hline
Formbook       & 400   \\ \hline
Qakbot         & 400   \\ \hline
Emotet         & 300   \\ \hline
AveMaria       & 290   \\ \hline
\end{tabular}
    \vspace{0.1cm}
\caption{\label{tab:malware}Malware families and corresponding sample count in dataset}
\end{table}

The malware families in our curated dataset are all significant threats and have seen continued
activity in recent years. Agent Tesla is a remote access trojan (RAT) malware to steal
credentials and user information and has been active for more than seven years~\cite{tesla}. New variants 
are continuously appearing within the growing number of attacks. Lokibot, another information-stealing
malware, has been observed by the Cybersecurity and Infrastructure Security Agency (CISA) to have
a considerable increase in popularity since July 2020~\cite{loki}. In 2021, Formbook malware
has been distributed via COVID-19 themed campaigns and has made it to the top of
Check Point Research's Global Threat Index~\cite{checkpoint-august}. Emotet, which is a
sophisticated Trojan that commonly acts as a downloader for other malwares, has also
appeared at the top of Check Point Research's index even after its global impact
had been reduced due to an international police operation~\cite{checkpoint-january}.
Lastly, the AveMaria RAT was observed to be distributed in a malicious email campaign and
disguised as Microsoft Word documents in the period of December 2020~\cite{avemaria}.
Hence there is no coincidence that these continuously active malwares were collected from MalwareBazaar's daily
uploads during the aforementioned period. Samples of these malware families made up the most
significant proportion of the daily malwares indicating highest rates of exposure and activity.

\subsubsection{Benign}
Our benign file samples are collected from two sources. One set of files is collected
from online repositories such as EXE Files~\cite{exefiles}. The other set consists of
standard files collected from Windows system directories.

\subsection{Experimental Process}
\label{subsec:exp-process}
In our experiments, we applied the 5-fold cross validation technique to obtain separate samples
for signature building and evaluation. Note that this is not applied to the dataset as a whole,
but instead needs to be performed separately for each malware family. The samples within each
family is randomly shuffled and split into five equal non-overlapping portions. One portion is used for testing.
The remaining samples are sent for clustering as in Section~\ref{subsec:clustering} and signature
building as in Section~\ref{subsec:sig} to obtain the sub-family signatures for each malware family.
The testing samples for all families are then combined into an aggregated testing set. Similarity
comparison is performed on each testing sample against the built signatures to obtain its
classification result. The same approach is also applied to the benign samples which is treated
as a single "family" with a set of detection signatures built.

Our experimental evaluation consists of two approaches:
\begin{enumerate}
    \item Binary classification approach
    \item Multi-class classification approach
\end{enumerate}

In binary classification approach, we focus on the proposed methodology's ability to only detect
a sample as either a malware or a benign file. For this, as long as a malware sample is matched
to a signature belonging to any of the malware families in our dataset, it is considered a correct
classification as it has been detected as malware regardless of the family.

In multi-class classification approach, the focus is on family attribution. Each test malware sample
must match to a sub-family signature belonging to the same malware family as its true label,
to constitute a true positive.

\subsection{Experimental Results}
\label{subsec:results}

\begin{figure}
  \centering{
    \includegraphics[scale=0.55]{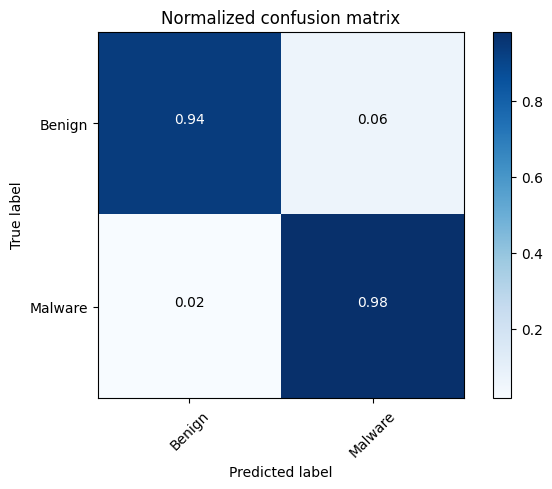}
    {\caption{Aggregated confusion matrix for binary classification}\label{fig:binary}}
  }
\end{figure}
\begin{figure}
  \centering{
    \includegraphics[scale=0.6]{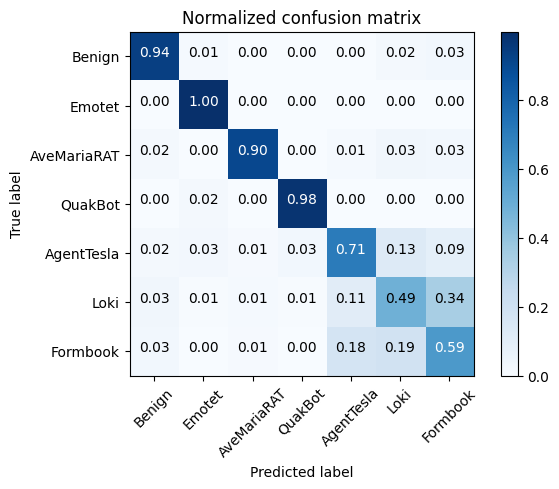}
    {\caption{Aggregated confusion matrix for multi-class classification}\label{fig:multi}}
  }
\end{figure}

Our experimental results are presented in Fig.~\ref{fig:binary} and Fig.~\ref{fig:multi} which are the
aggregated confusion matrices of the 5-fold cross validation for the binary classification and multi-class
classification approaches respectively.

For binary classification, we focus on two important performance metrics. The first metric is 
True Positive Rate (TPR), which is
the ratio of malware samples that can be correctly detected as malware. Following is the False
Positive Rate (FPR) which is the measure of benign samples being incorrectly classified as malwares.
In Fig.~\ref{fig:binary}, we can observe that the TPR is 98\% while the FPR is
at 6\%. With a high TPR of 98\%, the experiments indicate that the methodology has a high
confidence in being able to detect malware samples while only mistakingly classifying benign samples
as malwares at a low rate.

For multi-class classification, we also focus on the TPR of each malware family and
look at the rate of misclassification to each of other families to assess accurate 
malware family attribution. From Fig.~\ref{fig:multi}, we can observe high TPR of
100\%, 98\% and 90\% for the Emotet, Qakbot and AveMaria malware families respectively.
The performance for these families may be attributed to the fact that they contain distinct
sub-family clusters with a very high rate of similarity amongst samples within each cluster.
Again, this can be observed in Fig.~\ref{fig:quakbot-graphs} where little to no scattering of points
are observed for the marked clusters for the Qakbot malware family when \textit{eps} value of 0.01
is used as input to the DBSCAN algorithm. The ability to cluster with lower \textit{eps} values
indicates that points within clusters are very close to one another. Samples in the test set 
for these families are thus more easily matched to its family signatures instead of being misclassified
to the others leading to high TPR. This is an indication that there are very distinct and strong
opcode sequence patterns for the sub-family clusters in the above families.

The TPR rates for the other three families AgentTesla, Loki and Formbook are 71\%, 49\% and 59\%
respectively and are not as high, especially when compared to the previous families. This is an indication that the sub-family
clusters created and hence the signatures extracted for these families do not match the test
samples closely enough. Instead a higher similarity score was obtained for the signatures of other
malware families. However, a key observation can be made from the results in Fig.~\ref{fig:multi}.
A majority of the misclassifications for AgentTesla, Loki and Formbook actually occur amongst
themselves. Out of AgentTesla's misclassification rate of 29\%, a total of 22\% misclassification
was attributed to Loki and Formbook. 34\% of Loki's test samples were misclassified as Formbook and
11\% was misclassfied as AgentTesla while only 6\% was misclassified as benign and other families.
Similarly for Formbook, 18\% and 19\% test samples were misclassified as AgentTesla and Loki respectively,
out of 41\% total misclassification rate. In order to explain the above observation, we performed
further investigations to understand the root cause of these misclassifications.

\subsection{Investigations}
\label{subsec:investigations}

\begin{table*}[]
\centering
\begin{tabular}{|c|c|c|c|c|c|c|}
\hline
\textbf{Families} & Emotet & \multicolumn{1}{l|}{Qakbot} & \multicolumn{1}{l|}{AveMaria} & \multicolumn{1}{l|}{AgentTesla} & \multicolumn{1}{l|}{Loki} & \multicolumn{1}{l|}{Formbook} \\ \hline
Emotet            & -      & 0.795                       & 0.801                         & 0.784                           & 0.786                     & 0.782                         \\ \hline
Qakbot            & 0.795  & -                           & 0.840                         & 0.941                           & 0.927                     & 0.931                         \\ \hline
AveMaria          & 0.801  & 0.840                       & -                             & 0.886                           & 0.917                     & 0.901                         \\ \hline
AgentTesla        & 0.784  & 0.941                       & 0.886                         & -                               & 0.985                     & 0.987                         \\ \hline
Loki              & 0.786  & 0.927                       & 0.917                         & 0.985                           & -                         & 0.995                         \\ \hline
Formbook          & 0.782  & 0.931                       & 0.901                         & 0.987                           & 0.995                     & -                             \\ \hline
\end{tabular}
    \vspace{0.1cm}
    \caption{\label{tab:inves}Experimental investigation results of malware family similarities}
\end{table*}

Two forms of investigations were performed. The first is an experimental investigation to get an 
indication of the level of similarity each malware family as a whole has to one another. To do so,
we treated the samples within each family as a whole and created an opcode graph signature
to represent each family. This is similar to how signatures were built for the clusters in
Section~\ref{subsec:sig}. Similarity comparison was then performed between the signature
representing each family to that of all the other families.

Table~\ref{tab:inves} presents the investigation results. The table shows the score values
between each family and all other families in the dataset. In order for the results to be 
intuitive, we converted the scores to truely reflect similarity. This means that the higher the value,
the higher level of similarity. This is the opposite to the measure of distance used earlier in our
methodology introduction. To analyse the results, it would be important to gain a contextual view
of the values in the table. We can observe that the lowest score value in the table is 0.782,
which is between Emotet and Formbook. This tells that there should be considerable differences
between Emotet and Formbook in terms of opcode sequence patterns when comparing between malware
families. The score between Emotet and the rest of the families are relatively close in value,
indicating that Emotet is indeed very different to the rest. The highest scores in the table
belong to those in between AgentTesla, Loki and Formbook with values above 0.98. Specifically, 
Loki and Formbook are the most similar families producing the highest score value of 0.995.
The above helps to explain the multi-class classification results presented in Fig~\ref{fig:multi}.
As AgentTesla, Loki and Formbook have very similar opcode characteristics to one another, it
was natural that their samples would easily become recognized as belonging to any one of their
families. This lead to the high misclassification rate that was contained within these three families.

The other investigation performed was research-based to learn about the key functionalities of
each malware family and their similarities. Through our study, all the malware families are
forms of information stealers. However, a common property of AgentTesla, Loki and Formbook
malwares is that their main functions are to steal information and credentials from
browsers, mail clients and FTP clients as well as keylogging. While the other malware families
also steal data, they contain a variety of other functions. For example, Emotet 
is used as downloader for other malwares. Qakbot is a banking trojan and is able to drop additional
payload such as ransomware. Hence it is possible that the similar streamlined capabilities of
AgentTesla, Loki and Formbook lead to them having similar static features.

\subsection{Comparison with Existing Approach}
\label{subsec:comparison}

\begin{figure}
  \centering{
    \includegraphics[scale=0.6]{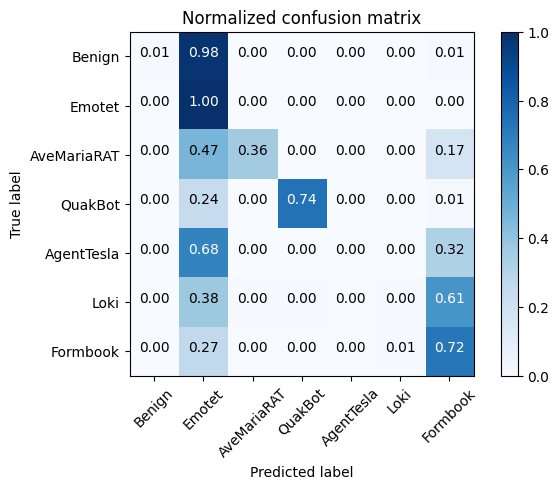}
    {\caption{Aggregated confusion matrix for multi-class classification using existing methodology without clustering}\label{fig:compare}}
  }
\end{figure}

In this section, we provide an experimental comparison of our methodology against the existing approach
in ~\cite{opcode-graph-2012}. The major difference in our methodology is the clustering process
described in ~\ref{subsec:clustering}. By performing experiments without the inclusion of the
clustering process, we can evaluate its impact on the performance and compare with the existing approach.

It is important to point out that we are unable to fully replicate the experimental procedure of
the work in ~\cite{opcode-graph-2012}. Firstly, the work uses only a representative sample from each
metamorphic family for comparisons to classify a new file. This was possible because based on their
investigations on the samples in their data, there was a clear separation of similarity score values
amongst samples in the same metamorphic family and that between the samples and benign files.
In other words, a boundary threshold value of the score could be set such that a score below the
threshold would indicate that a new sample is classified to the family, and a higher score means that the sample
does not belong to the family. However such a pattern does not exist in our dataset. Our
investigations indicate that samples within the same malware family do have considerable opcode
structural differences such that the scores would vary and a valid threshold is not possible.
For each malware family in the dataset, we thus combine the opcodes of all its samples to build
a single opcode graph signature which is the approach in ~\cite{co-opcode-graphs-2020} and also applied in our work.

We provide the multi-class classification results in Fig.~\ref{fig:compare}. A significant drop in 
performance can be observed when the clustering process is absent. The first observation that can be
made is close to total loss of ability to correctly classify benign files. This was an expected
observation as benign files should not bear much similarities with one another unless they
are related to the same application. Hence creating a single signature out of benign files would not
be meaningful as the signature would not be similar to any individual benign file. ~\cite{opcode-graph-2012}
uses only CYGWIN utilies as their benign files which would explain the ability to obtain a similar
match. 

The top performing malware families in our methodology had a reduction in TPR except for the Emotet
malware family. AgentTesla and Loki hit a 0\% TPR. Only Formbook had an improvement to its TPR.
An interesting observation is that most of the misclassifications for all classes went to the Emotet
malware family. However, such a phenomenom was not observed in the previous experimental results.
This indicates that without the clustering process, generalisation of the characteristics of each
malware family into a single signature is not effective. As each test sample does not bear significant
resemblance to its own family signature, it happens that the score is closest to Emotet for most cases.
Thus, the above experimental result highlights the importance of sub-family detection and
signature creation in our proposed methodology.

\subsection{Time Performance}
\label{subsec:time}
Our experiments were carried out on a machine with a processor model of 
Intel(R) Core(TM) i7-9700K CPU \@ 3.60GHz with eight cores available.
In order to perform the classification on an unknown sample, the first step is to perform
file disassembly to obtain opcodes and construct the opcode graphs. On average, this step takes
approximately 0.5 seconds. For opcode graph comparisons, a single comparison takes approximately
0.07 seconds. Hence the total time for comparison depends on the number of signatures
that is required for comparisons against. In our experiments, classification of each test sample
would take approximately 17 seconds with the utilisation of only one core of the processor. With the
implementation of multiprocessing, this can be significantly sped up if all available cores
are utilised. 

\vspace{-0.1cm}
\section{Conclusions}
\label{sec:conclusion}

In this paper, we presented an opcode-based methodology involving graph similarity comparison
for malware detection and malware family attribution. We integrated a clustering algorithm into
our approach to detect sub-family groups of malwares that potentially represent different versions
of the malware family. Opcode graph signatures are extracted from each cluster which are used for
similarity comparison to detect and attribute incoming samples. Experiments were performed
on a dataset containing different malware families to evaluate the proposed methodology. The method
achieved high recall and low false positives for binary classification. For multi-class classification,
high attribution performance was obtained for some malware families. For certain families
with relatively lower performance, investigations were performed to provide insights on the result.

There are several future work and enhancements that can be explored for the work. Further improvements
or new methodologies can be proposed for the opcode graph based similarity comparison techniques
that are used. Input parameters for the clustering algorithm resemble thresholds and are currently
manually determined. It may be beneficial to look into how to better set these parameters, possibly through
an automated approach. Another area is on scalability. The nature of signature comparison
approaches is that the more signatures there are, the processing time required increases linearly.
More signatures will be created from the addition of new malware samples and families and more
comparisons have to performed. Hence we should also look into efficiency improvements.

\bibliographystyle{plain}
\bibliography{references}
\end{document}